# Position Paper: Unlocking the Potential of AI Researchers in Scientific Discovery−What Is Missing?

Hengjie Yu
*School of Engineering*
*Westlake University*
Hangzhou, Zhejiang 310030, China
yuhengjie@westlake.edu.cn

Shuya Liu
*School of Engineering*
*Westlake University*
Hangzhou, Zhejiang 310030, China

Haiyun Yang
*School of Engineering*
*Westlake University*
Hangzhou, Zhejiang 310030, China

Yuping Yan
*School of Engineering*
*Westlake University*
Hangzhou, Zhejiang 310030, China

Maozhen Qu
*Department of Chemistry*
*National University of Singapore*
Singapore 117543, Singapore

Yaochu Jin
*School of Engineering*
*Westlake University*
Hangzhou, Zhejiang 310030, China
jinyaochu@westlake.edu.cn

***Abstract*—The potential of AI researchers in scientific discovery remains largely untapped. Over the past decade, AI for Science (AI4Science) publications in 145 Nature Index journals have increased fifteen-fold, yet they still account for less than 3% of the total publications. Drawing upon the Diffusion of Innovation theory, we project AI4Science's share of total publications to rise from 2.72% in 2024 to approximately 20% by 2050. Achieving this shift requires fully harnessing the potential of AI researchers, as nearly 95% of AI-driven research in these journals is led by experimental scientists. To facilitate this, we propose structured workflows and strategic interventions to position AI researchers at the forefront of scientific discovery. Specifically, we identify three critical pathways: equipping experimental scientists with accessible AI tools to amplify the impact of AI researchers, bridging cognitive and methodological gaps to enable more direct involvement in scientific discovery, and proactively fostering a thriving AI-driven scientific ecosystem. By addressing these challenges, we aim to empower AI researchers as key drivers of future scientific breakthroughs.**

***Keywords*—AI for Science, Scientific discovery, Diffusion of Innovation, Research paradigm, AI researcher**

## I. Introduction

Over the past decade, AI has become a powerful catalyst for scientific discovery [1], [2], [3]. Since 2015, its adoption and impact have expanded rapidly across scientific disciplines, driving unprecedented growth [4]. This surge has fueled the emergence of AI for Science (AI4Science), a field that leverages AI methods to tackle complex data challenges and uncover insights that were previously beyond reach. The impact of AI4Science is exemplified by the success of deep learning models in structural biology, chemistry, and biomedical research. For instance, AlphaFold [5] and ESMFold [6] have revolutionized protein structure prediction, dramatically improving the accuracy and efficiency of computational modeling compared to traditional physics-based methods. Similarly, in molecular representation learning, models like MolCLR [7] leverage contrastive learning to enhance the predictive power of molecular property predictions, facilitating drug discovery and materials science applications. Beyond structural and molecular biology, AI, especially large language models (LLMs), has also demonstrated remarkable utility in biomedical data analysis. GPTCelltype [8], an R package utilizing GPT-4, automates accurate cell type annotation from single-cell RNA sequencing data, greatly reducing the effort and expertise needed for this task. PathCha [9], a vision-language AI assistant tailored for human pathology, excels in diagnostic accuracy and user preference by integrating a specialized vision encoder with a pretrained LLM. AI has been recognized as a transformative technology and powerful paradigm in various scientific fields [3], such as health and medicine [10], materials [11], biology [12], chemistry [13], industry [14], and the environment [15].

Transformative technologies have periodically reshaped scientific discovery. Computational modeling and numerical simulation, such as density functional theory [16], revolutionized research in physics, chemistry, and engineering by enabling large-scale predictive simulations. The 1990s brought high-throughput experimental techniques like next-generation sequencing [17] and mass spectrometry [18], shifting biological science toward data-driven exploration. The early 2000s saw the rise of statistical learning and data science [19], enhancing pattern recognition and predictive modeling across disciplines. At the same time, AI is reaching a point of saturation in its traditional core areas, making it increasingly difficult for new researchers to find opportunities. Expanding AI into scientific discovery is therefore not only essential for accelerating progress across disciplines, but also offers a critical new frontier for AI researchers. In this sense, science needs AI to unlock complex discoveries, and AI researchers need science as a rich and sustainable direction for future innovation. With AI playing an expanding role in scientific discovery, a careful assessment of its current contributions and remaining challenges is both timely and necessary. This article explores two fundamental questions: To what extent has AI advanced scientific discovery, and what potential remains for further expansion and innovation? Additionally, what gaps still exist in the engagement of AI researchers with scientific discovery, and how can these be bridged?

To answer these questions, this research examines AI-related research papers published in leading natural and health science journals over the past decade. By analyzing these publications and their author affiliations, we aim to understand the current progress of AI4Science within high-impact research and identify pathways for AI researchers to engage in this field. Furthermore, we offer practical guidance and a structured workflow to support AI researchers in embarking on scientific discovery. By highlighting key entry points and methodological approaches, we aim to lower the barriers for AI researchers seeking to contribute to scientific advancements. Ultimately, we hope this survey serves as both an inspiration and a resource, facilitating broader integration of AI into scientific discovery while simultaneously expanding the frontiers of human knowledge, fostering mutual benefits for both AI researchers and the broader scientific community.

## II. Methods

### A. Data collection

The 145 Nature Index journals, selected by an expert panel of active scientists for their reputation and impact, have long been central venues for groundbreaking research in the natural and health sciences. We retrieved AI-related research articles published between 2015 and 2024 in the 145 journals covered by the Nature Index from the Web of Science Core Collection. The search query combined publication type (“Article”), publication years (2015-2024), and a comprehensive list of 50 AI-related terms in the topic field (e.g., “machine learning,” “deep learning,” “transformer model,” “reinforcement learning,” “computer vision,” etc.), yielding 20,603 articles. Retrieved metadata included titles, abstracts, publication years, author affiliations, and journal names. The retrieved data included information such as titles, abstracts, publication years, author affiliations, and journal names. To obtain the total number of research articles published in these journals during the same period, we used the search query (SO="Journal name" AND PY=2015-2024 AND DT=Article), which yielded a total of 1.16 million research articles.

### B. Research type classification

Although the search terms targeted AI-related studies, not all retrieved articles were classified as AI for Science, necessitating a classification of research types. We defined four categories of research: *Typical AI Research*, *AI for Science*, *Science for AI*, and *Not Applicable*. *Typical AI Research* refers to studies focused on the development or analysis of AI methods, such as machine learning, neural networks, deep learning, algorithms, or hardware, without direct application to natural or health sciences; *AI for Science* encompasses studies that apply AI techniques to address scientific challenges in fields such as materials science, chemistry, physics, biology, and medicine, where AI is explicitly used to enhance discovery, prediction, data analysis, or scientific understanding; *Science for AI* denotes research that leverages findings from natural or cognitive sciences to inspire or improve AI models, architectures, or hardware, with the goal of advancing AI itself rather than solving scientific problems; *Not Applicable* covers studies outside these domains, including those unrelated to AI, those applying AI in non-scientific contexts, or those where AI is mentioned only peripherally.

Given the scale of the dataset, manual classification was impractical. Therefore, we employed a hybrid workflow that combines automated reasoning models with human verification (Fig. 1). Large and small reasoning models equipped with carefully designed prompts, a voting mechanism, and optimization of search terms, prompts, and model usage were integrated to ensure classification accuracy. The process begins with three state-of-the-art 20-32B reasoning models, Qwen3-32B [20], GPT-OSS-20B, and DeepSeek R1-32B [21], released in 2025. Articles (title and abstract) unanimously classified as *AI for Science* by all three models were directly assigned. In cases of disagreement, a larger 120B model (GPT-OSS-120B) was introduced. If the large model’s output aligned with the majority decision of the smaller models, the article was categorized accordingly. To address LLMs’ discrepancies in classification, human verification was applied. A random sample of 400 LLM classifications was assessed for accuracy, leading to four rounds of optimization, which resulted in a classification accuracy exceeding 95%.

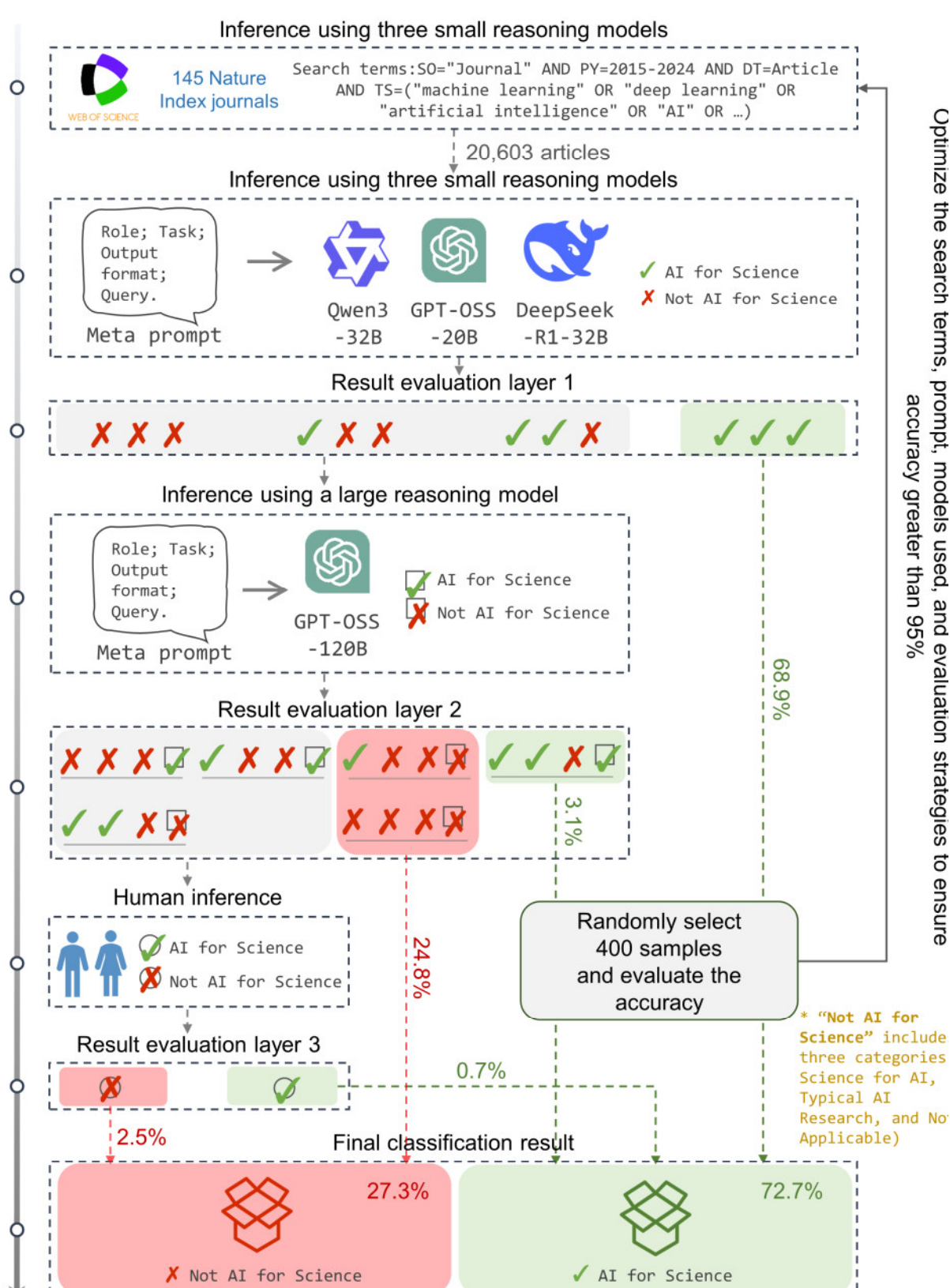

Fig.1. Workflow for classifying research articles in collaboration with reasoning models and human oversight.

### C. Author affiliation classification

To classify author affiliations, we applied the similar hybrid workflow used in Research Type Classification, utilizing both large and small reasoning models. Affiliations were categorized into three groups: *AI Institute*, *Science Institute*, and *Not Applicable*. *AI Institute* includes institutions explicitly focused on AI, machine learning, or related fields. *Science Institute* refers to institutions dedicated to research in natural or health sciences, like biology, chemistry, or medicine, without reference to AI. *Not Applicable* includes non-research organizations, comprehensive research institutions, or institutions where AI or scientific research is not the primary focus.

### D. Scientific entity extraction and curation

We used GPT-OSS-120B with carefully designed prompts to extract terms from article titles and abstracts. The extracted terms were processed through a multi-step curation pipeline: first, inconsistent delimiters were corrected and malformed or abnormally short entries removed; then, terms were split into individual entities, stripped of whitespace and punctuation, and standardized to first-letter capitalization. To reduce redundancy, we normalized terms by ignoring case, punctuation, and hyphens, and clustered those with shared prefixes. Finally, we filtered terms by length and word count, grouped similar expressions, and ranked the clusters by total frequency for downstream analysis.

### E. Scientific entity classification and word cloud plotting

Using the LLM-human collaboration workflow, each identified term was classified into one of three categories based on a structured prompt: *AI* (terms directly related to artificial intelligence technologies, such as machine learning, neural networks, and deep learning), *Science* (terms associated with natural or health sciences, such as physics, biology, or medicine), or *Not Applicable* (general or unrelated terms not specific to either domain). High-frequency terms underwent minor manual refinement, such as merging obvious synonyms and standardizing capitalization, to improve clarity and consistency for visualization. We then used the *wordcloud* package to generate a word cloud, clearly depicting the most frequently mentioned entities in the dataset.

## III. Results And Discussion

### A. AI4Science: A Rapidly Growing, Yet Relatively Minor Component of Scientific Discovery

We compiled a dataset of 20,603 AI-related research articles published in these journals over the past decade. Although our search terms were designed to capture AI-related studies, not all retrieved articles qualified as "*AI for Science*." Using a classification framework based on a hybrid system of large and small reasoning models, augmented with carefully designed prompts, and human verification, we ultimately categorized 72.7% of the articles as "*AI for Science*."

Further analysis reveals that the number of AI-related articles exhibits a rapid increase over the last ten years, with both the absolute number and proportion rising across 145 Nature Index journals (Fig. 2A). The growth in AI-related research is especially evident in recent years, with the number of such articles in 2024 being fifteen times higher than in 2015, and the relative proportion of AI-related articles has also surged, reaching twelve times the level observed in 2015. This significant rise highlights the escalating adoption of AI methodologies across various scientific disciplines. Notably, 21 journals each published over 50 AI-related articles in 2024 alone, indicating a broad integration of AI. Furthermore, 16 journals had more than 5% of their total publications focused on AI, with five surpassing 9%. Among these, *Nature Methods* stands out, with AI-related articles comprised 15.87% of its total output in 2024, reflecting the pivotal role of AI in advancing specific fields. However, they still represent only 2.72% of total publications, indicating that AI is still in the early stages of widespread adoption in science.

The Diffusion of Innovation theory [22], developed by Everett Rogers, describes how new ideas and technologies spread through a population over time. It posits that adoption follows an S-curve, with innovations initially embraced by a small group of innovators and early adopters, followed by the early majority, late majority, and finally laggards. Based on this pattern, we anticipate that AI-related research will continue to grow rapidly over the next few decades, with its expansion gradually slowing as it reaches maturity (Fig. 2B). As the innovation matures, its adoption rate increases rapidly, then slows as it becomes widely accepted and integrated into everyday practices. AI's application in scientific discovery is likely following a similar pattern. Early adopters in fields like computational biology, materials science, and data-driven research have led the charge, demonstrating the transformative potential of AI. As more researchers integrate AI into their work and the technology becomes more accessible, adoption will spread to the early majority and late majority, solidifying its place as a central tool in scientific research. By 2050, we project that AI-related research will stabilize at around 20% of the total Nature Index journal publications, reflecting AI's fully integrated role in modern research.

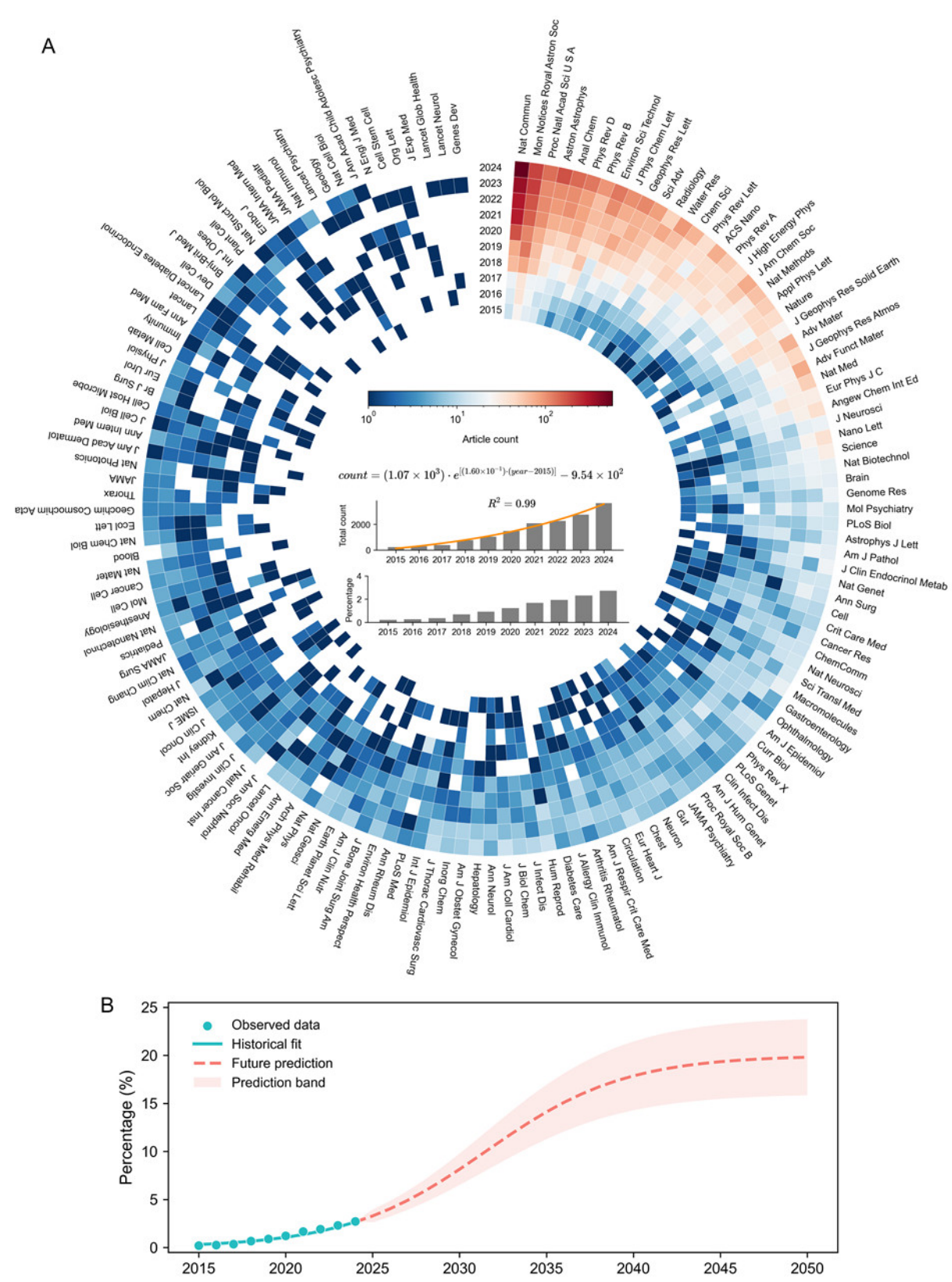

Fig.2. Publication and projected growth trends of AI-related research articles in 145 Nature Index journals. (A) Publication trends of 20,603 AI-related research articles in Nature Index journals (2015-2024). (B) Projected growth trend of AI-related research in Nature Index journals based on a logistic growth model. The shaded area represents the upper and lower bounds of this 20% projection.

### B. AI Researchers Increasingly Contribute to Scientific Discovery in a Supporting Role

To elucidate the roles of AI researchers in the current AI for Science wave, we analyzed 61,759 author affiliations from 20,603 AI-related articles. These affiliations were categorized into three groups: *AI Institute*, *Science Institute*, or *Not Applicable* (for those not clearly affiliated with either AI or Science). The trends in author affiliations are illustrated in Fig. 3. AI institutions are increasingly involved in scientific discovery, yet they largely maintain a supportive role despite their growing significance. The average number of AI institutions per paper has risen from 0.16 in 2015 to 0.38 in 2024 (Fig. 3A). In contrast, the average number of institutions per research article and the total number of scientific institutions have remained relatively stable, experiencing a decline between 2018 and 2022, followed by a slight upward trend in recent years. The proportion of papers with AI institutions as author affiliations has grown from 13.33% to 25.30% (Fig. 3B). Additionally, AI institute-led research saw a notable rise in 2016, followed by a gradual and uneven upward trend in the subsequent years (Fig. 3C), with the average percentage over the past three years at 5.66%. Similarly, the average ranking of the first-listed AI institution has improved from 6.97 in the first year to 3.68 over the past nine years. However, scientific institutions continue to dominate AI-driven scientific discovery, with the average ranking of the first-listed science institution remaining stable at around 1.1, shifting slightly from 1.09 in the first three years to 1.12 in the latest three years (Fig. 3D).

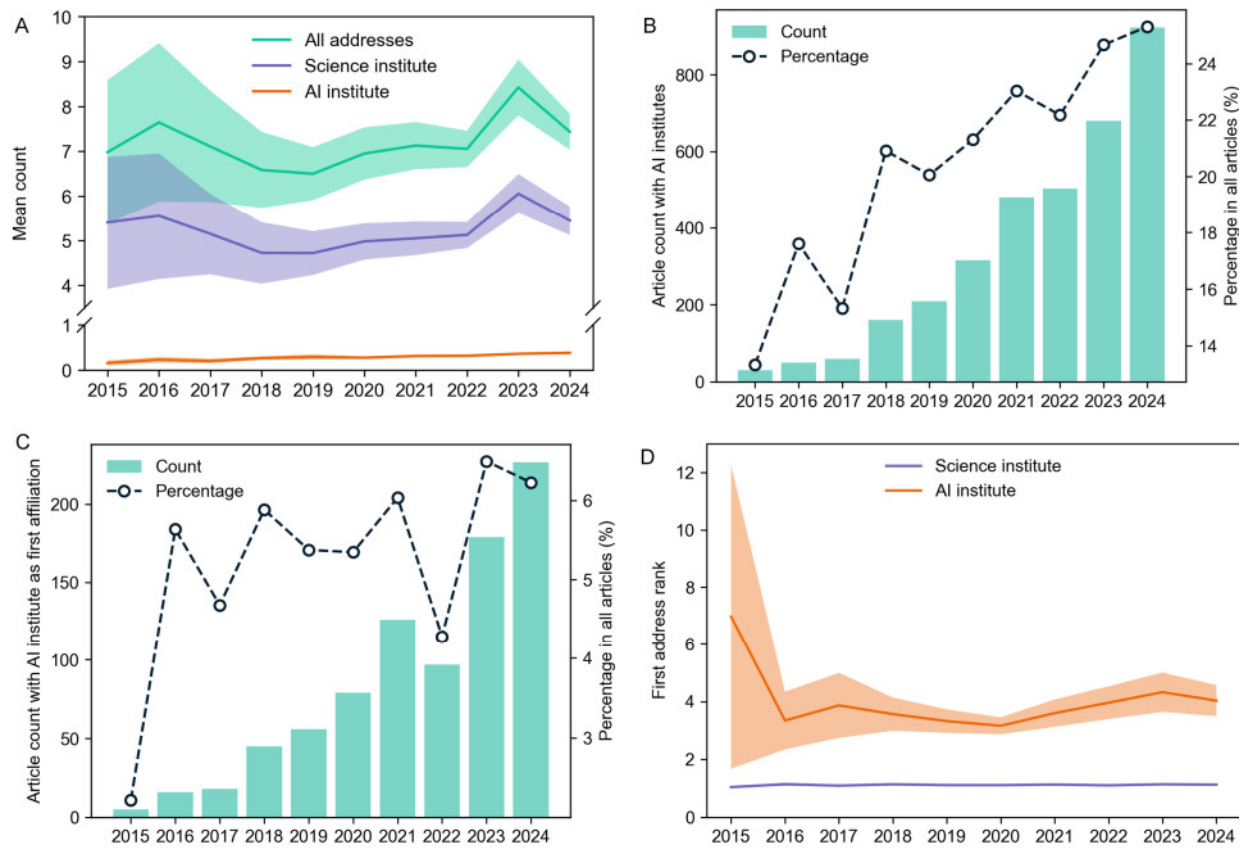

Fig.3. Trends in author affiliations of 20,603 AI-related research articles (2015-2024). (A) Number of institutions per article by institution type. (B) Number and proportion of articles involving AI-related institutions. (C) Number and proportion of articles with AI-related institutions as the first affiliation. (D) Rank of the first-affiliated AI and Science institution. The shaded area represents the 95% confidence interval.

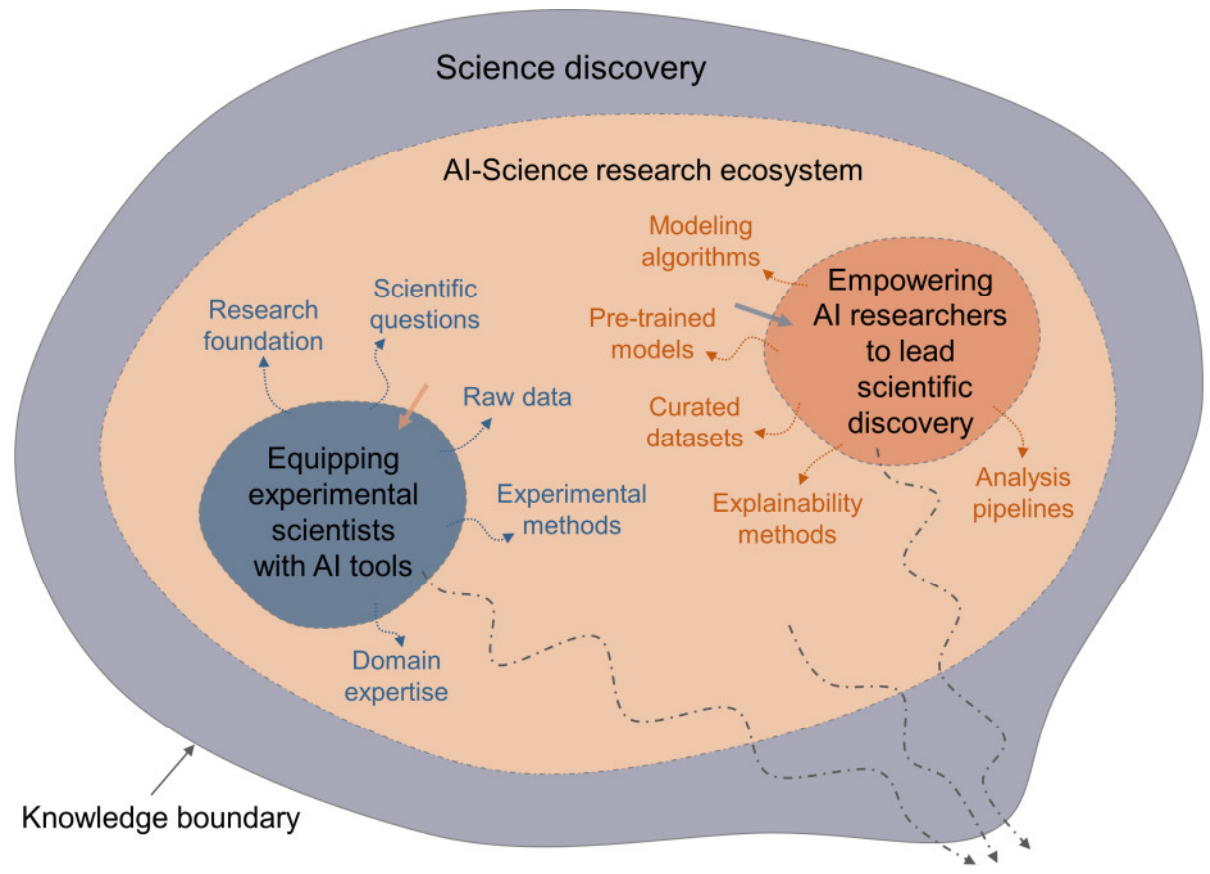

Fig.4. Schematic representation of three key strategies for unlocking the potential of AI researchers in scientific discovery: (1) equipping experimental scientists with user-friendly AI tools, (2) enabling AI researchers to take a more direct and active role in scientific discovery, and (3) fostering a thriving AI-driven scientific ecosystem to sustain long-term innovation.

Given that many Nature Index journals may be less familiar to AI researchers, we also analyzed participation in five prominent interdisciplinary journals widely recognized in the AI community: Nature, Nature Communications, Science, Science Advances, and PNAS. These five journals account for 3,262 (21.7%) of all AI-for-Science articles. AI engagement here has consistently outpaced the broader Nature Index: in 2015, AI institutions appeared more frequently, as co-authors and first affiliations, than the overall average by roughly twofold. By 2024, 36.3% of articles included an AI affiliation, though only 10.7% listed one as the first. Notably, AI participation plateaued after 2021, suggesting a shift from rapid expansion to consolidation in these venues.

### *C. Unlocking The Potential of AI Researchers in Scientific Discovery: Three Key Directions*

AI is rapidly accelerating scientific discovery. However, despite the widespread use of AI tools by experimental scientists, the role of AI researchers remains underexplored (Fig. 3). Achieving the projected growth, from 2.72% to 20%, in AI-driven scientific contributions requires stronger engagement from AI researchers. We identify three key directions, which are both ongoing and yet to be fully defined, that are essential to realizing this transformation, as illustrated in Fig. 4.

Developing user-friendly AI tools for scientific discovery offers major opportunities for AI researchers by increasing the real-world impact of their work. Providing experimental scientists with AI tools for data collection [23], modeling and analysis [24], and experiment design and conduction [25] can significantly accelerate scientific discovery. across a range of scientific fields. In fact, early adopters among experimental scientists have already begun utilizing AI tools in their research. As shown in Fig. 3B and Fig. 3C, 75% of AI-related articles are currently independently authored by experimental scientists. The next key step in advancing the application of AI is the development of more user-friendly tools and platforms, which will be discussed below. This is crucial for further driving the adoption of AI among experimental scientists.

The success of AI4Science depends not only on experimental scientists using AI tools but also on AI researchers actively participating in scientific discovery. By combining AI expertise with domain knowledge, these researchers can develop specialized methods to address scientific challenges. Although the number of AI researchers leading AI-driven discoveries has increased, this growth has slowed and remains limited (Fig. 3C). We attribute this limitation to two main gaps: a cognitive gap (regarding what scientific discoveries AI can be applied to) and a methodological gap (in terms of the workflow in which AI researchers lead scientific discoveries). Discussions on bridging these gaps will be presented in later sections.

Collaboration is central to AI4Science and extends beyond individual projects to the broader research ecosystem. A representative example is SHapley Additive exPlanations (SHAP) [26], [27], an interpretability method developed by AI researchers and widely adopted across many scientific fields. For instance, SHAP has been used to quantify the contributions of individual metabolites to disease risk, identifying key metabolites influencing the risk of 24 investigated diseases [28]. It has also been applied to interpret nanomaterial-plant-environment interactions [29], [30], providing insights into the factors affecting the root uptake of metal-oxide nanoparticles and their interactions within the soil. Additionally, SHAP was employed to analyze the contributions of raw ingredients and their interactions, offering valuable insights into the factors influencing the compressive strength of alkali-activated materials [31]. Its widespread use is largely enabled by an accessible open-source package, making the method easy for experimental scientists to apply. Importantly, experimental scientists often value model interpretability more than small performance gains, as explanations help generate new scientific hypotheses. Building such collaborative ecosystems, supported by shared tools and standards, can accelerate AI4Science by improving transparency, reproducibility, and adoption.

### *D. Four Approaches to Equipping Experimental Scientists with AI*

AI will not replace scientists because scientific discovery requires human intuition, creativity, and the ability to navigate uncertainty—capabilities that AI lacks [32], [33]. However, scientists who leverage AI can process vast amounts of data, automate complex analyses, and generate insights more efficiently, giving them a significant advantage in accelerating breakthroughs and expanding the frontiers of knowledge [34], [35]. Here, we discuss four key approaches, both established and emerging, that enable researchers to harness AI effectively, lowering barriers to adoption while accelerating scientific discovery (Fig. 5). These approaches not only illustrate how experimental scientists can harness AI tools but also present significant opportunities for AI and software researchers. Building user-friendly platforms of this kind, akin to existing statistical analysis software but enhanced with AI capabilities, can bridge the gap between advanced machine learning and practical scientific applications, enabling broader adoption and innovation.

User-friendly AI platforms can streamline data-driven research by automating model selection, training, and model explanation (Fig. 5A). Scientists only need to provide data, define prediction targets, and set resource constraints, after which the platform generates models, analyses, and interpretable reports. Model explanation and the resulting reports are critical components as they directly contribute to scientific understanding, which is one of the primary objectives of scientific research [36], [37]. Besides, many scientific fields lack sufficient experimental data, yet AI can still provide meaningful insights by integrating domain knowledge with LLMs (Fig. 5B). Although LLMs face challenges such as hallucinations and limited reasoning, retrieval-augmented generation can improve accuracy by incorporating external knowledge sources [38]. By building domain-specific knowledge bases, these platforms enable zero- or few-shot predictions, allowing scientists to explore hypotheses even with limited data. As experimental scientists increasingly apply AI in both data-rich [39], [40] and data-scarce [41], [42] settings, user-friendly platforms will further promote adoption and enhance the role of AI in scientific discovery.

High-quality data is fundamental to machine learning-driven scientific discovery [43], yet much of this data remains embedded in unstructured scientific literature. Text, figures, and supplementary materials are not readily accessible to existing informatics systems, making data extraction a major bottleneck for AI-enabled science [44]. Consequently, converting published literature into usable, structured datasets is critical for advancing AI-assisted scientific discovery. Although LLMs have improved literature-based data extraction [45], their ability to handle complex textual content, figures, and supplementary files remains limited. With the emergence of more advanced LLMs, particularly those fine-tuned for data extraction tasks [46], and the rapid development of multimodal LLMs, substantial improvements in extraction accuracy and coverage are expected. In this workflow, researchers specify target articles and variables of interest, after which AI systems automatically retrieve, curate, and structure relevant data, enabling more efficient meta-analyses, comparative studies, and data-driven hypothesis generation (Fig. 5C).

Beyond data extraction, AI can also support scientific discovery by autonomously designing and executing experiments. (Fig. 5D). Researchers define hypotheses, control variables, and desired outputs, while an AI-powered platform, enhanced by knowledge-augmented reasoning, develops an optimized experimental plan. Robotics then execute experiments, and a validation module assesses the reliability and quality of the data. Although AI- and robot-driven automated laboratories are still in their early stages, initial applications have already emerged [47], [48], demonstrating remarkable potential to transform scientific research [49]. This approach minimizes manual labor, increases reproducibility, and enables high-throughput hypothesis testing, ultimately accelerating scientific progress.

### *E. Scientific Domain "Ocean" with AI Application Potential*

AI offers immense potential in scientific discovery, yet many opportunities remain underexplored by AI researchers. Currently, the application of AI in science is largely concentrated in fields with abundant structured datasets, such as foundational life sciences and medical research. Prominent examples include protein structure and function prediction, single-cell annotation, and drug discovery (Fig. 6A). The success of AI in these areas is largely driven by the availability of well-curated, large-scale datasets and clearly defined machine learning tasks. However, scientific discovery extends far beyond these domains, and numerous critical challenges in other fields remain largely untapped by AI researchers.

AI has strong potential to transform a wide range of scientific fields, including materials design, material-bio interactions, chemical and biosynthesis, environmental remediation, climate modeling, and industrial optimization (Fig. 6B). These fields involve complex, multi-dimensional problems that require AI for data integration, prediction, and interpretation. However, many of these domains remain underexplored by the AI community, largely due to limited data availability and the need for specialized domain knowledge. To understand where AI is currently being applied in science, we analyzed AI-related research articles and identified frequently occurring scientific terms (Fig. 6C). This analysis highlights the scientific areas where experimental researchers are already using AI to address core research questions. Expanding AI's impact in less-explored domains will require developing methods that can handle complex and unstructured scientific data, as well as AI-driven approaches for data acquisition, experiment design, and real-time analysis to improve research efficiency and accelerate discovery.

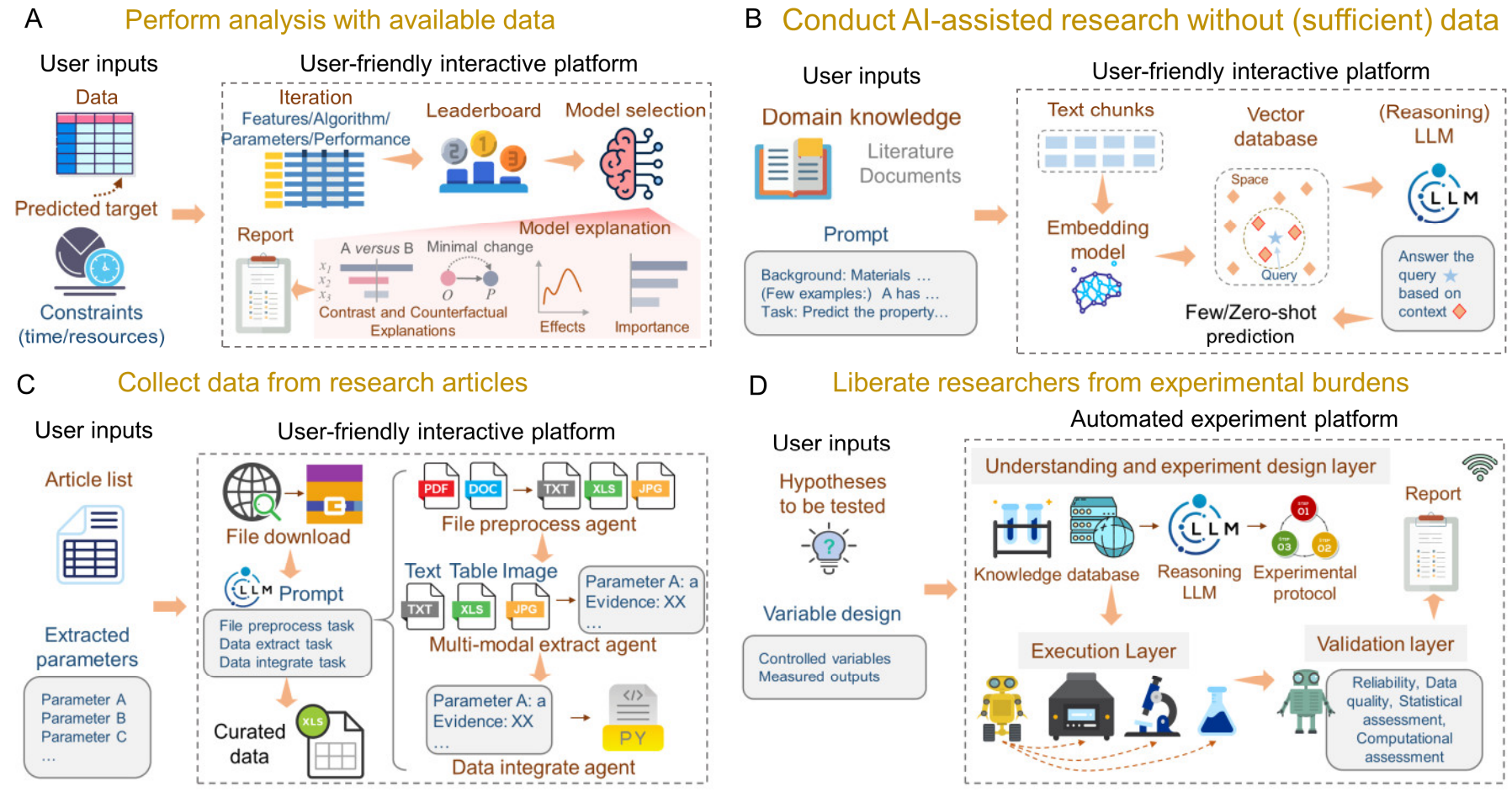

Fig.5. Schematic diagram of four approaches to equipping experimental scientists with AI. (A) User-friendly platforms facilitate modeling and explanation when researchers have access to sufficient data. (B) Domain-enhanced LLMs enable zero-shot and few-shot inference in data-scarce scenarios. (C) The integration of LLMs, multimodal systems, and multi-task frameworks makes automated data extraction from literature increasingly feasible. (D) Advances in LLMs and robotics are driving the development of autonomous experimental platforms, freeing researchers from labor-intensive experiments.

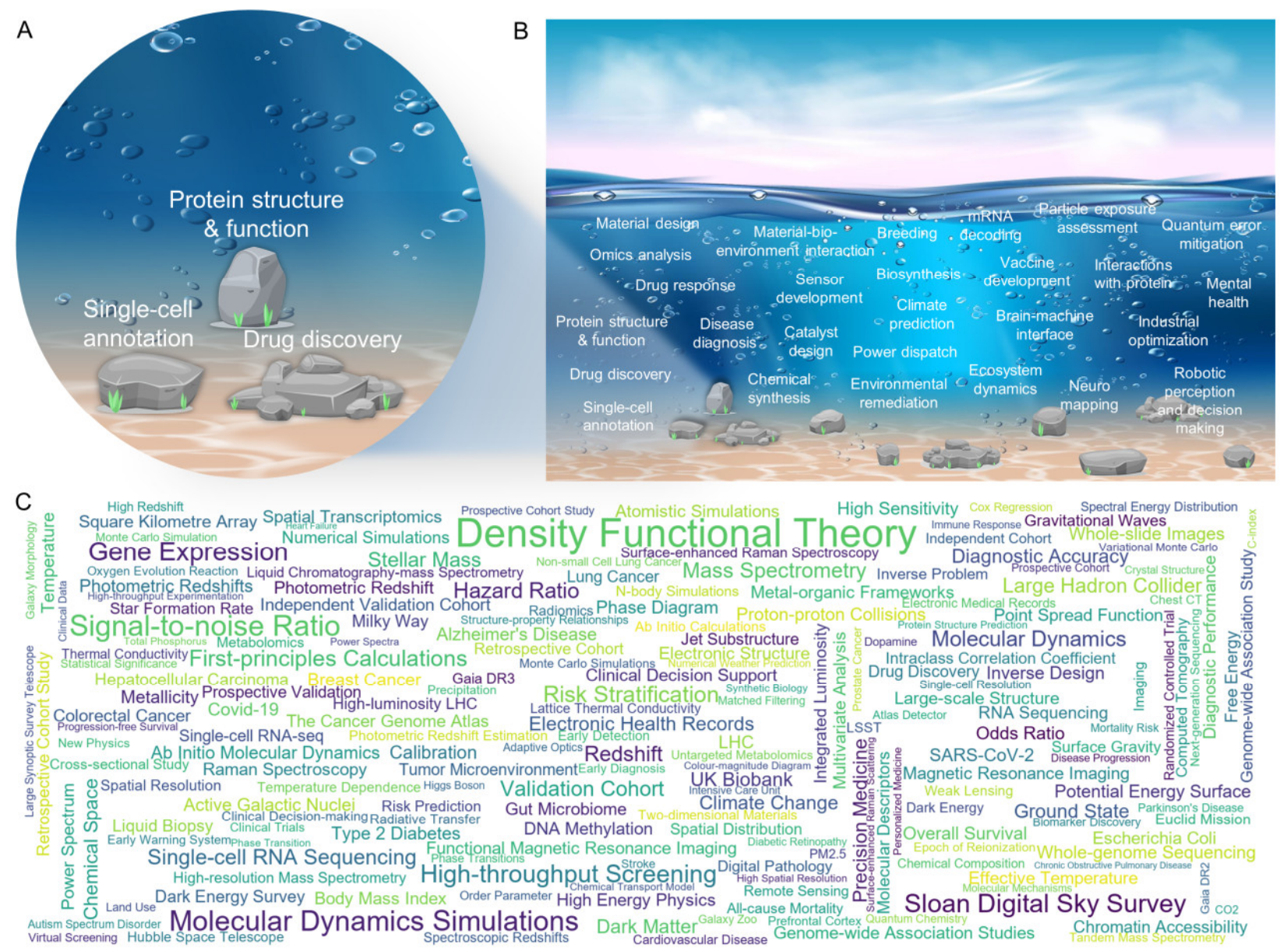

Fig.6. AI involvement and application potential across scientific domains. (A) Current hot topics in AI-driven science, mainly supported by abundant structured data and well-defined learning tasks, such as protein modeling and drug discovery. (B) Scientific domains with broad future potential for AI, highlighting complex, multidisciplinary problems that require new AI methods for unstructured data and knowledge integration. (C) Frequently occurring scientific terms in AI-related publications, reflecting research areas where AI is already actively applied to fundamental scientific problems.

### *F. AI4Science Workflow for AI Researchers: From Tool Providers to Key Contributors*

There are many AI4Science workflows for scientific discovery in various fields, such as proteomics [50], materials [39], biomedicine[51], and medical imaging [52], but most are designed for researchers in specific fields rather than for AI researchers. We propose an AI4Science workflow for AI researchers (Fig. 7).

The first and most crucial step is to identify suitable scientific problems, which can be guided by the researchers' interests, experience, and available resources. Areas where AI has already been applied may be a good starting point (Fig. 6C), as experimental scientists may have used basic machine learning methods to address such problems. Understanding the chosen scientific domain is a prerequisite for conducting efficient research, and with the development of LLM-based agents [53], this will become easier. For example, LLMs have enabled the automated generation of novel scientific ideas [54] and scientific hypotheses [55]. Moreover, researcher agents based on LLMs have been developed to aid scientific discovery. For instance, Google DeepMind's AlphaEvolve combines the creativity of an LLM with algorithms that refine and filter mathematical and computer science solutions [56]. This system allows for both innovative idea generation and solution improvement. The Virtual Lab, an LLM-PI-led, human-in-the-loop agent team, used an ESM-AlphaFold-Rosetta pipeline to design 92 nanobodies, with experiments validating functional binders, including two improved against JN.1/KP.3 variants [57].

Overall, AI plays three main roles across different fields: prediction, comprehension, and innovation (discovery or design). These roles are logically connected. For example, in the case of proteins, one can first predict the structure or functional annotation of a protein, then understand the relationship between the sequence or structure and function, and finally, based on this understanding, design new proteins to achieve specific functions. In data-rich fields, one can skip the understanding step, but in many cases, understanding is closely tied to scientific discovery.

Formal experiments can generally be divided into three steps: data collection, modeling and analysis, and experimental validation. For AI researchers, we do not recommend generating data through experiments or simulations due to the associated complexity. Instead, it is more practical to focus on fields with existing datasets or those where data can be extracted from the literature, such as the nanobio interface [58]. In data-rich fields, the competition is more intense. In many domains, high-quality public data are scarce, and literature-based data extraction can provide a competitive advantage, especially with the help of LLMs and AI tools. Modeling and analysis are core strengths of AI researchers, so we do not elaborate further here, but we emphasize the importance of interpretability and causal inference, as scientific understanding remains a central goal of research [36]. Finally, experimental validation is essential for confirming computational findings. This step can often be achieved through standardized protocols, collaborations with experimental labs, shared platforms, or commercial research services. Looking ahead, automated laboratories powered by LLMs and robotics are expected to further lower the barrier to experimental validation [49], [59], making it increasingly accessible.

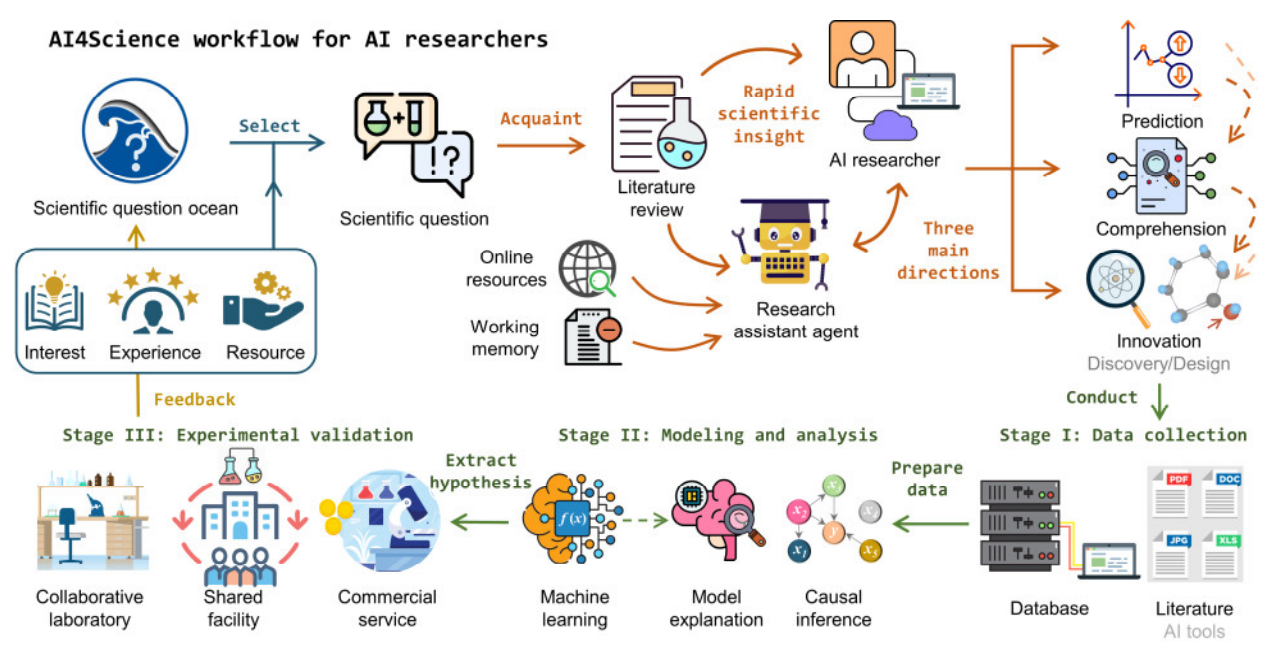

Fig. 7 AI4Science workflow for AI researchers.

## IV. Conclusion

AI4Science has grown rapidly over the past decade, yet its full potential, particularly the role of AI researchers, remains underutilized. Although AI is increasingly integrated into scientific research, its application is still largely led by experimental scientists, with AI researchers often serving in supporting roles. This limits the depth of AI's impact on discovery. To accelerate progress, it is essential to position AI researchers as active contributors rather than mere tool developers. By fostering deeper collaboration and bridging cognitive and methodological gaps, they can drive more transformative advances in science. Looking ahead, the future of AI4Science depends on a well-defined human-AI collaboration paradigm that leverages AI's analytical strengths while keeping human researchers in control of scientific reasoning and validation. Addressing challenges such as hallucinations and defining clear boundaries for AI use will be critical. Moreover, experimental validation remains the ultimate safeguard of AI-driven discoveries, ensuring scientific rigor across disciplines. With these shifts, AI4Science could expand from its current 2.72% share of Nature Index publications to 20% by 2050, unlocking new frontiers in scientific discovery.

## Acknowledgment

This work is supported by Westlake Education Foundation (Grant No. 103110846022301) and China Postdoctoral Science Foundation (Grant No. 2024M762941). The authors acknowledge the use of ChatGPT solely for language polishing and assume full accountability for the manuscript's accuracy and integrity.